\documentclass[preprint,prd,nofootinbib,tightenlines,amsmath,showpacs]{revtex4}
\usepackage{enumerate}
\usepackage{multirow}
\usepackage[utf8]{inputenc}
\usepackage{anysize}
\usepackage{textcomp}
\usepackage{epsfig}
\usepackage{graphics,color} 
\usepackage{verbatim}
\usepackage{afterpage}
\usepackage{amsmath}    
\usepackage{soul}
\usepackage{xcolor,cancel}
\DeclareUnicodeCharacter{00A0}{ }

\usepackage{blindtext}

\catcode`\@=11
\def\sla@#1#2#3#4#5{{%
 \setbox\z@\hbox{$\m@th#4#5$}%
 \setbox\tw@\hbox{$\m@th#4#1$}%
 \dimen4\wd\ifdim\wd\z@<\wd\tw@\tw@\else\z@\fi
 \dimen@\ht\tw@
 \advance\dimen@-\dp\tw@ \advance\dimen@-\ht\z@
 \advance\dimen@\dp\z@
 \divide\dimen@\tw@ \advance\dimen@-#3\ht\tw@
 \advance\dimen@-#3\dp\tw@ \dimen@ii#2\wd\z@
 \raise-\dimen@\hbox to\dimen4{%
 \hss\kern\dimen@ii\box\tw@\kern-\dimen@ii\hss}%
 \llap{\hbox to\dimen4{\hss\box\z@\hss}}}}

\def\cpto{\mathrel {\vcenter {\baselineskip 0pt \kern 0pt
    \hbox{$H_{r.f.}$} \kern 0pt \hbox{$\longrightarrow$} }}}
\def\slashed#1{%
 \expandafter\ifx\csname sla@\string#1\endcsname\relax
{\mathpalette{\sla@/00}{#1}}
\fi}
\def\declareslashed#1#2#3#4#5{%
 \expandafter\def\csname sla@\string#5\endcsname{%
#1{\mathpalette{\sla@{#2}{#3}{#4}}{#5}}}}
 \catcode`\@=12
\declareslashed{}{/}{.08}{0}{D}
 \declareslashed{}{/}{.1}{0}{A}
 \declareslashed{}{/}{0}{-.05}{k}
 \declareslashed{}{/}{.1}{0}{\partial}
 \declareslashed{}{\not}{-.6}{0}{f}

\def\lsim{\mathrel {\vcenter {\baselineskip 0pt \kern 0pt
    \hbox{$<$} \kern 0pt \hbox{$\sim$} }}}
\def\gsim{\mathrel {\vcenter {\baselineskip 0pt \kern 0pt
    \hbox{$>$} \kern 0pt \hbox{$\sim$} }}}

\newcommand{\bea}{\begin{eqnarray}}
\newcommand{\eea}{\end{eqnarray}}

\begin{document}

\baselineskip=15pt
\preprint{CoEPP-MN-15-11}

\title{ T-odd correlations from the top-quark chromoelectric dipole moments in lepton plus jets top-pair events}

\author{Alper Hayreter$^{1}$\footnote{Electronic address: alper.hayreter@ozyegin.edu.tr} and German Valencia$^{2}$\footnote{Electronic address: German.Valencia@monash.edu }}


\affiliation{$^{1}$ Department of Natural and Mathematical Sciences, Ozyegin University, 34794 Istanbul Turkey.}

\affiliation{$^{2}$ School of Physics and Astronomy, Monash University, 3800 Melbourne Australia.\footnote{On leave from Department of Physics, Iowa State University, Ames, IA 50011.}}

\date{\today}

\vskip 1cm
\begin{abstract}

There exist several recent studies of the top-quark chromoelectric dipole moment (CEDM) in the context of searching for CP violating signals in top-quark pair production at the LHC. Most of these studies constrain the CEDM either from deviations in the top-pair cross section from its standard model value, or from T-odd asymmetries in the dimuon channel. Motivated by ATLAS and CMS interest, we extend the study of T-odd asymmetries to the lepton plus jets channel. At the parton  level, using {\tt MADGRAPH5}, we identify the most promising signals and their statistical sensitivity. We find that the signals with larger sensitivity to the CEDM require distinguishing between $b$ and $\bar{b}$ jets and propose a simple way to address this.

\end{abstract}


\maketitle

\section{Introduction}

As the LHC becomes a top-quark factory it becomes increasingly interesting to search for new sources of CP violation at high energy. In the absence of a compelling new model with CP violation affecting the top quark, a good first step is to study the lowest-dimension operators that can induce the desired effect. In particular, since the dominant production process for the top quark at the LHC is gluon fusion, the top-quark CEDM becomes the benchmark for this type of studies. 

The flavor diagonal dipole couplings between top quarks and gluons of magnetic and electric type are conventionally written as
\begin{eqnarray}
{\cal L}=\frac{g_s}{2}\ \bar{t}\ T^a\sigma^{\mu\nu}\left(a_t^g+i\gamma_5 d_t^g\right) \ t\ G^a_{\mu\nu}.
\label{defcoup}
\end{eqnarray}
with $d_t^g$ being the CP-odd CEDM. As written, this operator is not gauge invariant under the full standard model  gauge group. Within the context of effective field theories and assuming that the particle discovered at the LHC is the standard model (SM) Higgs boson, the gauge invariant generalization of Eq.~(\ref{defcoup}) is, in the notation of Ref.~\cite{Buchmuller:1985jz},
\begin{eqnarray}
{\cal L} = g_s\frac{d_{tG}}{\Lambda^2}\ \bar{q}_3\sigma^{\mu\nu}T^a t\  \tilde\phi G^a_{\mu\nu}  +\ {\rm H.c.}
\label{GIanocoup}
\end{eqnarray}
where $q_3$ is the third generation SM quark doublet, $\phi$ is the scalar doublet, $\tilde\phi_i =\epsilon_{ij}\phi_j$ and the $SU(3)$ generators are normalized as ${\rm Tr}(T_aT_b)=\delta_{a,b}/2$. Electroweak symmetry is spontaneously broken when the scalar acquires a vacuum expectation value $<\phi>=v/\sqrt{2}$, $v\approx 246$~GeV resulting in the correspondence
\begin{eqnarray}
d_{t}^g &=& \frac{\sqrt{2} \ v}{\Lambda^2}{\rm ~Im}(d_{tG}).
\label{anomcouplings}
\end{eqnarray}
The main effect of imposing invariance under the SM group in this case is that the  top-quark CEDM also modifies Higgs production in association with a top-quark pair~\cite{DeRujula:1990db,Degrande:2012gr,Hayreter:2013kba,Choudhury:2012np,Ellis:2014dva,Bramante:2014gda}. For our numerical estimates we will implicitly assume a new physics scale $\Lambda=1$~TeV, but rescaling to any other value can be simply read off Eq.~(\ref{GIanocoup}).

The top-quark CEDM has been studied at length in connection with top-quark pair production and decay~\cite{Bernreuther:1992be,Brandenburg:1992be,Atwood:1992vj,Bernreuther:1993hq,Cheung:1995nt,Choi:1997ie,Sjolin:2003ah,Martinez:2007qf,Antipin:2008zx,Gupta:2009wu,Gupta:2009eq,Choudhury:2009wd,Hioki:2009hm,HIOKI:2011xx,Kamenik:2011dk,Ibrahim:2011im,Hioki:2012vn,Biswal:2012dr,Baumgart:2012ay,Hioki:2013hva,Bernreuther:2013aga,Kiers:2014uqa,Englert:2014oea,Rindani:2015vya,Gaitan:2015aia,Bernreuther:2015yna}. 
A typical result is that of Ref.~\cite{Gupta:2009wu} where it is found that using CP-odd observables the 5$\sigma$ statistical sensitivity with 10~fb$^{-1}$ at 14 TeV to $d^g_t$ is of order $0.1/m_t$. 

Most of the previous studies rely on measuring deviations from the SM cross section, or on measuring T-odd observables in the dimuon channel. Our purpose in this paper is to extend the study of CP-odd spin correlations written directly as T-odd triple products of momenta in the lab frame \cite{Donoghue:1987ax} to the lepton plus jets channel for the top-quark pair.

\section{Numerical study}

For our numerical study we generate multiple event samples for the process $pp \to t \bar{t} \to b\bar{b} \ell^\pm \nu jj$ where $\ell =\mu,e$ at 14~TeV center-of-mass energy that we summarize in the Appendix.  The CEDM coupling is implemented in  {\tt MADGRAPH5} \cite{MadGraph} with the aid of {\tt FEYNRULES} \cite{Christensen:2008py}.\footnote{The code is available from the authors upon request.} We use the resulting UFO model files to generate events for several values of $d_{tG}$ in a range motivated by our previous results from Ref.~\cite{Gupta:2009wu}. 

To single out  CP violating couplings, we consider  T-odd correlations that involve the beam, $t$ quark, $b$ quark, lepton and jet momenta with as many as eight momentum factors.  These include all the ones that have been previously discussed in the literature to our knowledge. For each of these correlations, ${\cal O}_i$ we quantify our bounds using the integrated asymmetry in the lab frame defined by
\begin{eqnarray}
{\cal A}_i&=&\frac{\sigma({\cal O}_i>0)-\sigma({\cal O}_i<0)}{\sigma_{SM}}.
\label{Tasym}
\end{eqnarray}
The events preserve all spin correlations between production and decay of the top quarks as  the full amplitude for the process is calculated. In each case we generate event samples with $10^6$ events {\it after} cuts, implying a $1\sigma$ statistical sensitivity to all asymmetries at the $(\sigma/\sigma_{SM}\times 0.1)\%$ level. We have used the following kinematic cuts
\begin{eqnarray}
&&|{p_T}_{\mu,j}|>20 ~~,~~ |{p_T}_{b,\bar{b}}|>25,\nonumber \\
&&|\eta_{\mu,b,\bar{b},j}|<2.5 ~~,~~ \slashed{E}_T>30,\nonumber \\
&&\Delta R_{ik} > 0.4~~ (i,k=\mu,b,j)
\label{eqcuts}
\end{eqnarray}
The beam momenta are written as $P^\mu = p_1^\mu +p_2^\mu$ and $q^\mu= p_1^\mu -p_2^\mu$. All observables involving $q$ will be quadratic in $q$ respecting the $p_1\leftrightarrow p_2$ identical particle symmetry of the initial state.

We begin with the point ${\rm Im}(d_{tG})=3$ for $\Lambda=1$~TeV, which corresponds to almost twice the $5\sigma$ sensitivity for 10~fb$^{-1}$ found in Ref.~\cite{Gupta:2009wu} in the dilepton channel. We use the dilepton channel as a benchmark to calibrate the size of the asymmetries, so we start by repeating the calculation for this case and tabulating our results in the second column of Table~\ref{tab0} for 13 different observables (the notation is further explained in the Appendix). In the third column of Table~\ref{tab0}, we present the corresponding correlations for the lepton plus jets channel for the idealized case in which the $d$ quark (or $s$ quark) could be tracked. Of course, these are not observable, but they allow us to compare directly with the dilepton channel since  the charged lepton or down-type quark are the best spin analyzers in top decay. The results in Table~\ref{tab0} indicate that, with appropriate modifications, almost all the observables in the dilepton channel can be reproduced (at least in principle) in the lepton plus jets channel. In many cases the correspondence involves keeping track of the lepton charge as a way to keep the momenta in the correlation in the right order. There are three correlations that do not involve non-$b$ jet momenta, $A_{2,5,12}$, and they agree for both channels within statistical error (see note on $A_5$ below).

\begin{table}[htp]
\begin{tabular}{|l|c|c|}
\hline &$pp\to t\bar{t} \to b\bar{b} \ell^+\ell^- \slashed{E}_T $ & $pp\to t\bar{t} \to b\bar{b} \ell^\pm jj \slashed{E}_T $ \\ \hline
${\cal{O}}_1$ &  $ \epsilon(t,\bar{t},\ell^+,\ell^-)$ &  $q_\ell\ \epsilon(t,\bar{t},\ell,d)$ \\
$A_1$  & -0.1540&   $-0.1535 \xrightarrow[]{p_t\to p_{t-vis}}-0.1114$\\
\hline 
${\cal{O}}_2$ &  $\epsilon(t,\bar{t},b,\bar{b})$ &  $\epsilon(t,\bar{t},b,\bar{b})$ \\
$A_2$  & -0.0358 &  $-0.0311 \xrightarrow[]{p_t\to p_{t-vis}}-0.0527$\\
\hline 
${\cal{O}}_3$ &  $\epsilon(b,\bar{b},\ell^+,\ell^-)$ &  $q_\ell\ \epsilon(b,\bar{b},\ell,d)$\\
$A_3$  & -0.0902 &  -0.0838\\
\hline 
${\cal{O}}_4$ &$\epsilon(b^+,b^-,\ell^+,\ell^-)$  & $\epsilon(b^\ell,b^d,\ell,d)$\\
$A_4$  &-0.0340 &    -0.0319 \\
\hline 
${\cal{O}}_5$ & $q\cdot(\ell^+-\ell^-) \epsilon(b,\bar{b},\ell^++\ell^-,q)$ & $q_\ell q\cdot\ell \epsilon(b,\bar{b},\ell,q)$ \\
$A_5$  &  -0.0309 &  -0.0115 \\
\hline 
${\cal{O}}_6$ &$\epsilon(P,b-\bar{b},\ell^+,\ell^-)$  & $q_\ell\ \epsilon(P,b-\bar{b},\ell,d)$ \\
$A_6$  &   0.0763&    0.0742 \\
\hline 
${\cal{O}}_7$ &  $q\cdot(t-\bar{t})\epsilon(P,q,\ell^+,\ell^-)$& $q_\ell\ q\cdot(t-\bar{t}) \epsilon(P,q,\ell,d)$ \\
$A_7$  &  -0.0373  &$-0.0325 \xrightarrow[]{p_t\to p_{t-vis}}-0.0257$\\
\hline 
${\cal{O}}_8$ &$q\cdot(t-\bar{t})(P\cdot\ell^+\epsilon(q,b,\bar{b},\ell^-)+P\cdot\ell^-\epsilon(q,b,\bar{b},\ell^+))$& $ q\cdot(t-\bar{t})(P\cdot\ell\epsilon(q,b,\bar{b},d)+P\cdot d\epsilon(q,b,\bar{b},\ell))$\\
$A_8$  & 0.0074& $0.0113 \xrightarrow[]{p_t\to p_{t-vis}}0.0094$ \\
\hline 
${\cal{O}}_9$ & $q\cdot(\ell^+-\ell^-)\epsilon(b+\bar{b},q,\ell^+,\ell^-)$ &$ q\cdot \ell \epsilon(b+\bar{b},q,\ell,d)$\\
$A_9$  & 0.0089 & 0.0051\\
\hline 
${\cal{O}}_{10}$ &$q\cdot(b-\bar{b})\epsilon(b,\bar{b},q,\ell^++\ell^-)$&$ q\cdot(b-\bar{b})\epsilon(b,\bar{b},q,d)$\\
$A_{10}$  & -0.0069&-0.0045\\
\hline 
${\cal{O}}_{11}$ &$q\cdot(b-\bar{b})\epsilon(P,q,b+\bar{b},\ell^+-\ell^-)$&$q_\ell q\cdot(b-\bar{b})\epsilon(P,q,b+\bar{b},d)$ \\
$A_{11}$  & -0.0147 & 0.0140  \\
\hline 
${\cal{O}}_{12}$ &  $q\cdot(b-\bar{b})\ \epsilon(P,q,b,\bar{b})$& $q\cdot(b-\bar{b})\ \epsilon(P,q,b,\bar{b})$\\
$A_{12}$  &0.0058 & 0.0041\\
\hline 
${\cal{O}}_{13}$ &$\epsilon(P,b+\bar{b},\ell^+,\ell^-)$&  $q_\ell\epsilon(P,b+\bar{b},\ell,d)$ \\
$A_{13}$ & 0.0032 & 0.0025    \\
\hline 
${\cal{O}}_{14}$ &-&  $\epsilon(P,b+\bar{b},\ell,d)$ \\
$A_{14}$ & - & -0.0013   \\
\hline 
\end{tabular}
\caption{Comparison of asymmetries in the dilepton and semileptonic channels for $d_{tG}=3$, $\Lambda =1$~TeV. The latter do not yet correspond to observable asymmetries and serve only for this comparison.}
\label{tab0}
\end{table}

Several comments with respect to Table~\ref{tab0} are pertinent:
\begin{itemize}
\item Identifying the T-odd correlations in semileptonic top-pair decay with the corresponding one in dilepton decay by using $d\leftrightarrow \ell$ results in the same asymmetry within statistical error. This is as expected and an important check at this stage of the calculation.
\item When the identification is not exact, as in $A_5$ there is a small difference. We can check this is the case by repeating the semileptonic asymmetry with the correlation $q_\ell q\cdot(\ell -d) \epsilon(b,\bar{b},(\ell+d),q)$, in which case we find  $A_5 = -0.0284$.  
\item The momentum of the top quark that decays leptonically cannot be fully reconstructed. To see what effect this has, we repeat the calculation of the semileptonic asymmetries that involve a top-quark momenta, replacing it with the visible top-quark momenta defined as
\begin{equation}
p_{t-vis} = p_b+p_\ell
\label{ptvis}
\end{equation}
This results in the asymmetries also shown in Table~\ref{tab0}. It is not necessary to repeat this exercise for the dilepton events since we do not concern ourselves with those in this paper, and this was done in Ref.~\cite{Gupta:2009wu}.
\item The operator ${\cal O}_{13}$ is T-odd but CP even, and as such it cannot be generated by a CEDM in agreement with our numerical result. It can, however, be used to look for unitarity phases. In the lepton plus jets channel it is possible to construct a T and CP odd operator with the same momenta, ${\cal O}_{14}$.
\end{itemize}

In Table~\ref{tab2} we construct the observables in terms of jet momenta, but still at the parton level. To do this we must define the non-$b$ jet in a CP blind way and there is more than one definition that works. We illustrate the effect of four different definitions in  Table~\ref{tab2}:
\begin{itemize}
\item $j_1$ the jet in this case is the hardest non-$b$ jet (largest $p_T$).
\item $j_2$ the jet in this case is the second hardest non-$b$ jet. Note that at the parton level this is the same as the softest non-$b$ jet.
\item $j_3$ the jet is the one closest to the  $b$ jet in the hadronic top decay side of the process, as determined by  $\Delta R$.
\item $j_4$ the jet reconstructs the  $W$ that decays hadronically, that is, $p_j = p_U + p_D$ where $U=u {\rm ~or~}c$ and $D=d{\rm ~or~}s$.
\end{itemize}
In addition, we replace any top-quark momenta with the visible top-quark momenta as defined above.

\begin{figure}[h]
\includegraphics[scale=0.4]{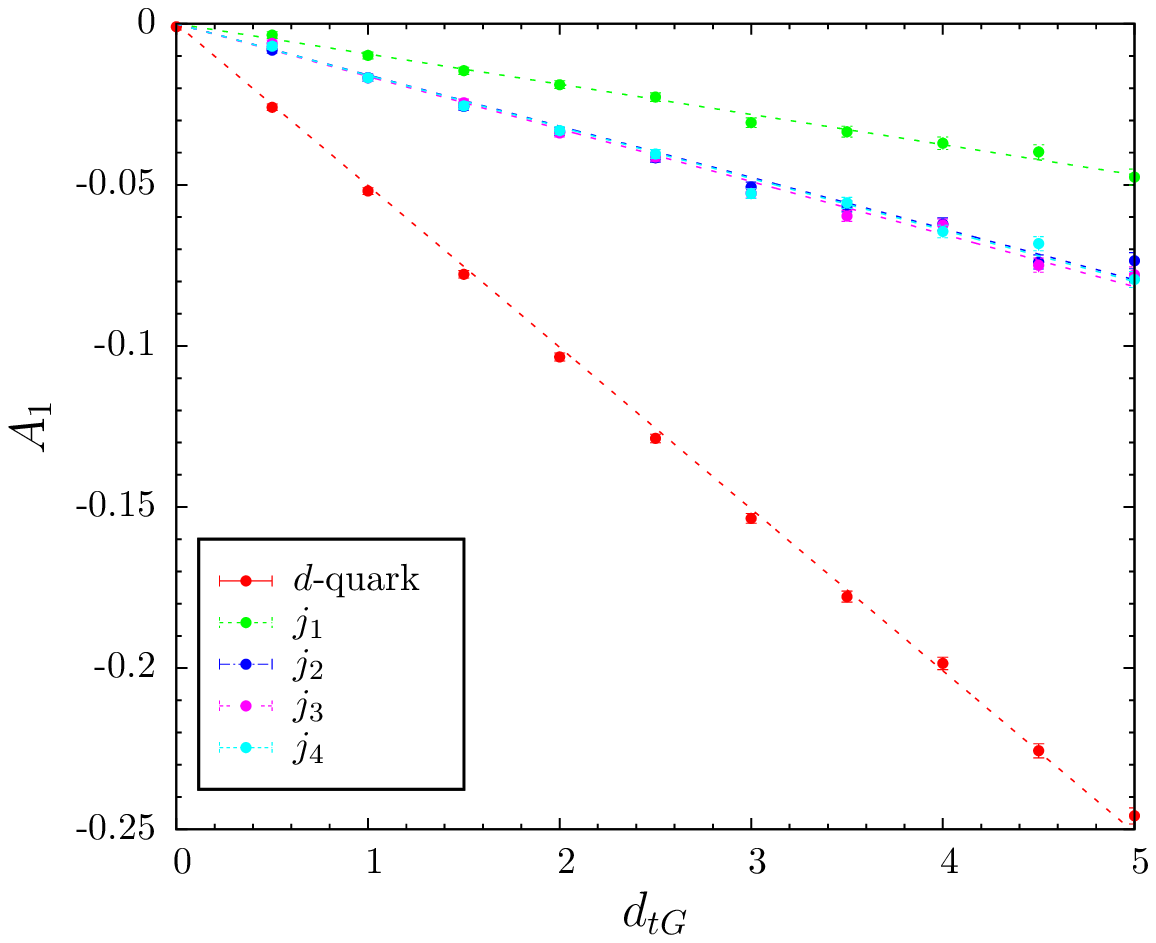} \hspace*{0.1cm}
\includegraphics[scale=0.4]{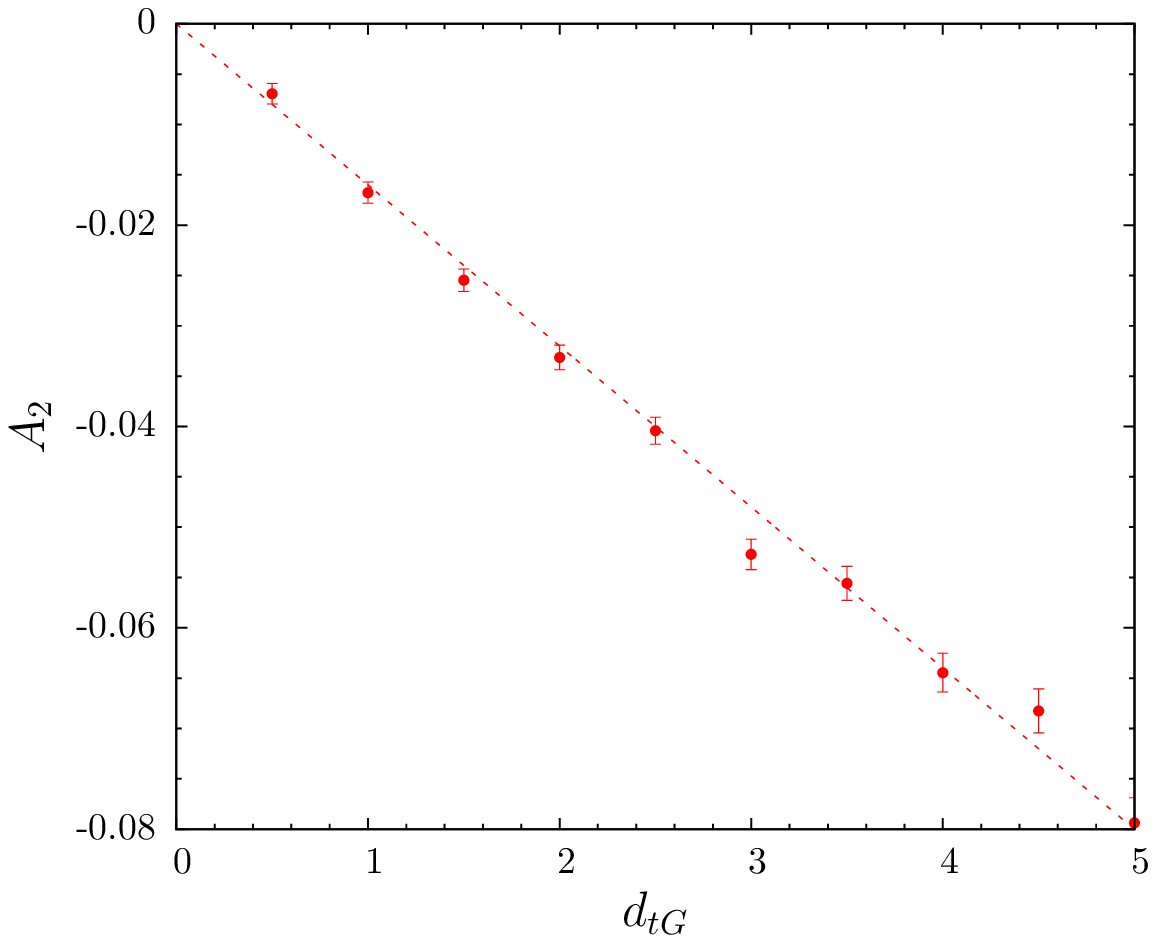}  \hspace*{0.1cm}
\includegraphics[scale=0.4]{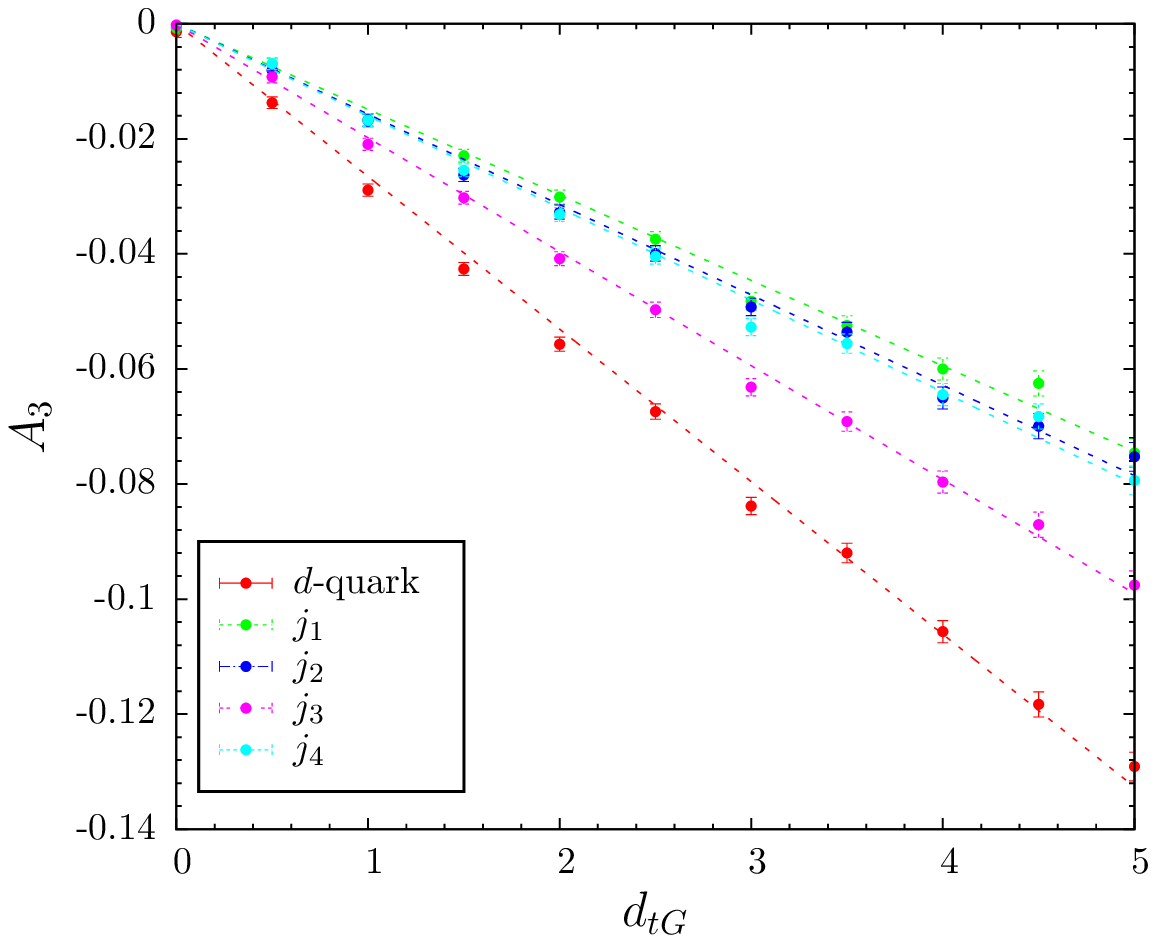}
\includegraphics[scale=0.4]{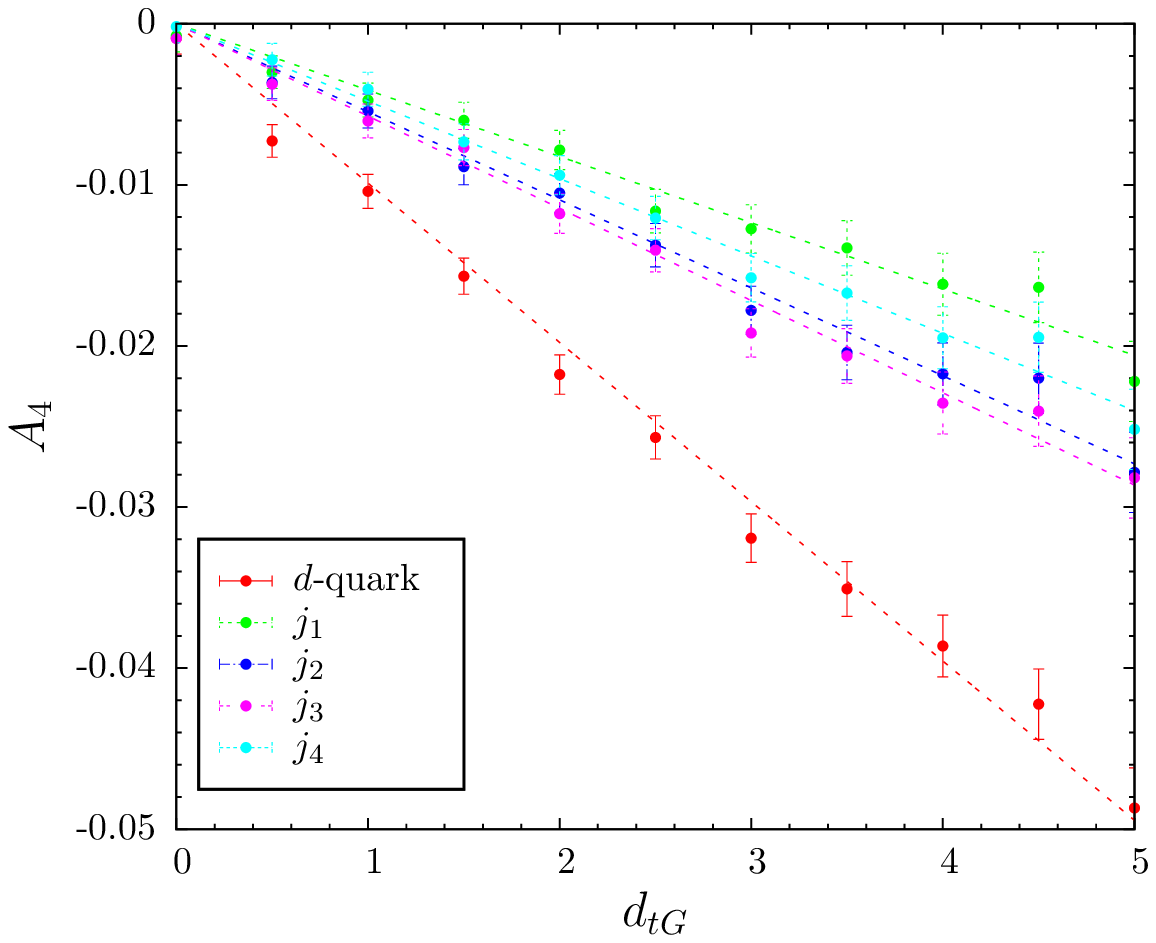}  \hspace*{0.1cm}
\includegraphics[scale=0.4]{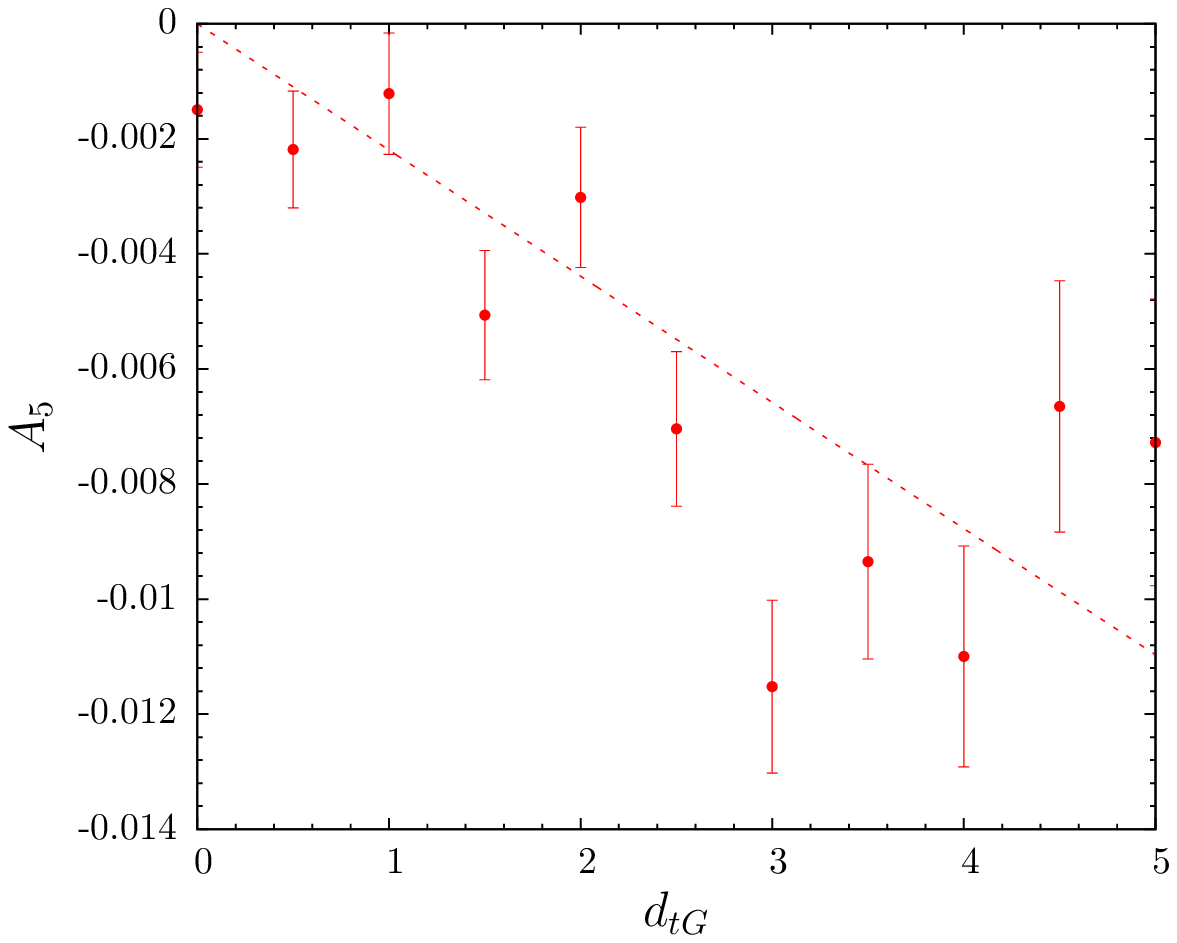}  \hspace*{0.1cm}
\includegraphics[scale=0.4]{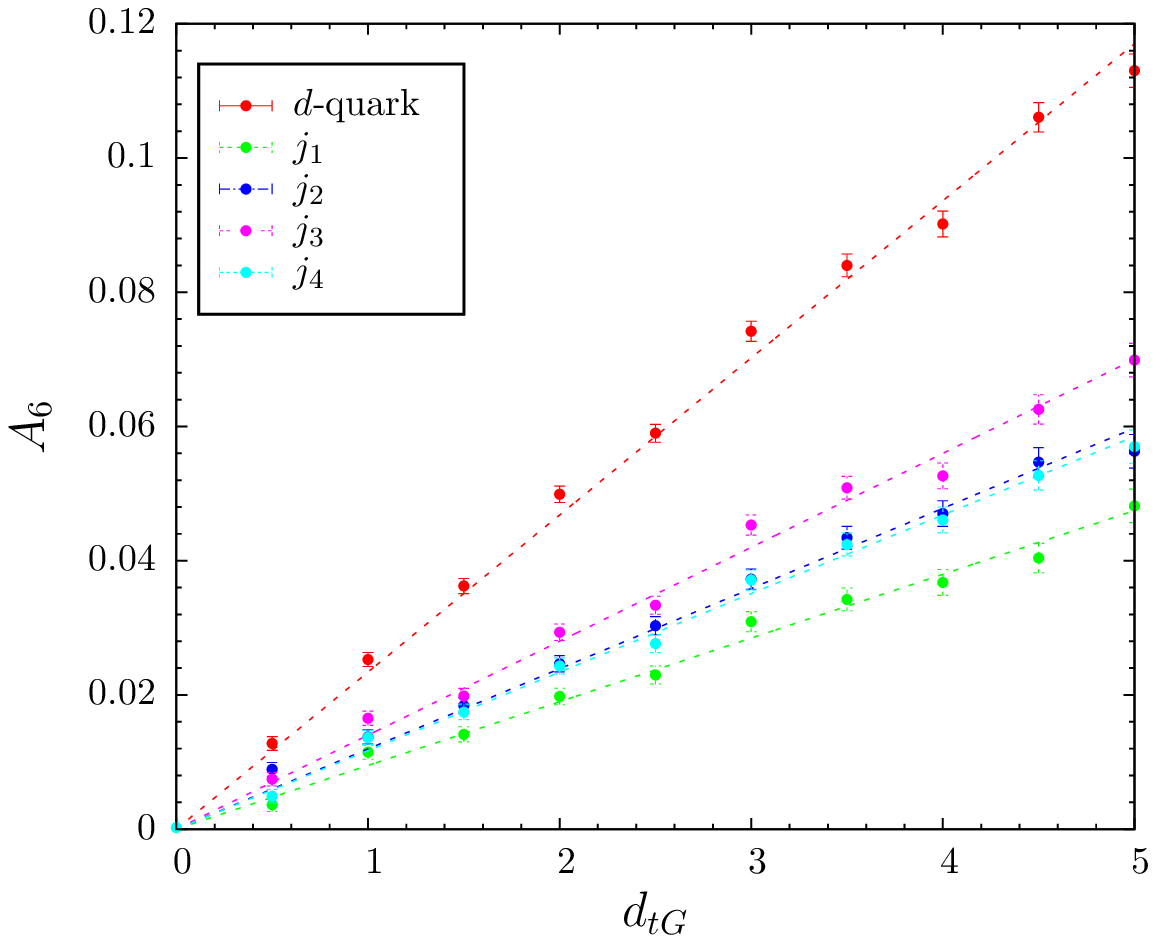} 
\includegraphics[scale=0.4]{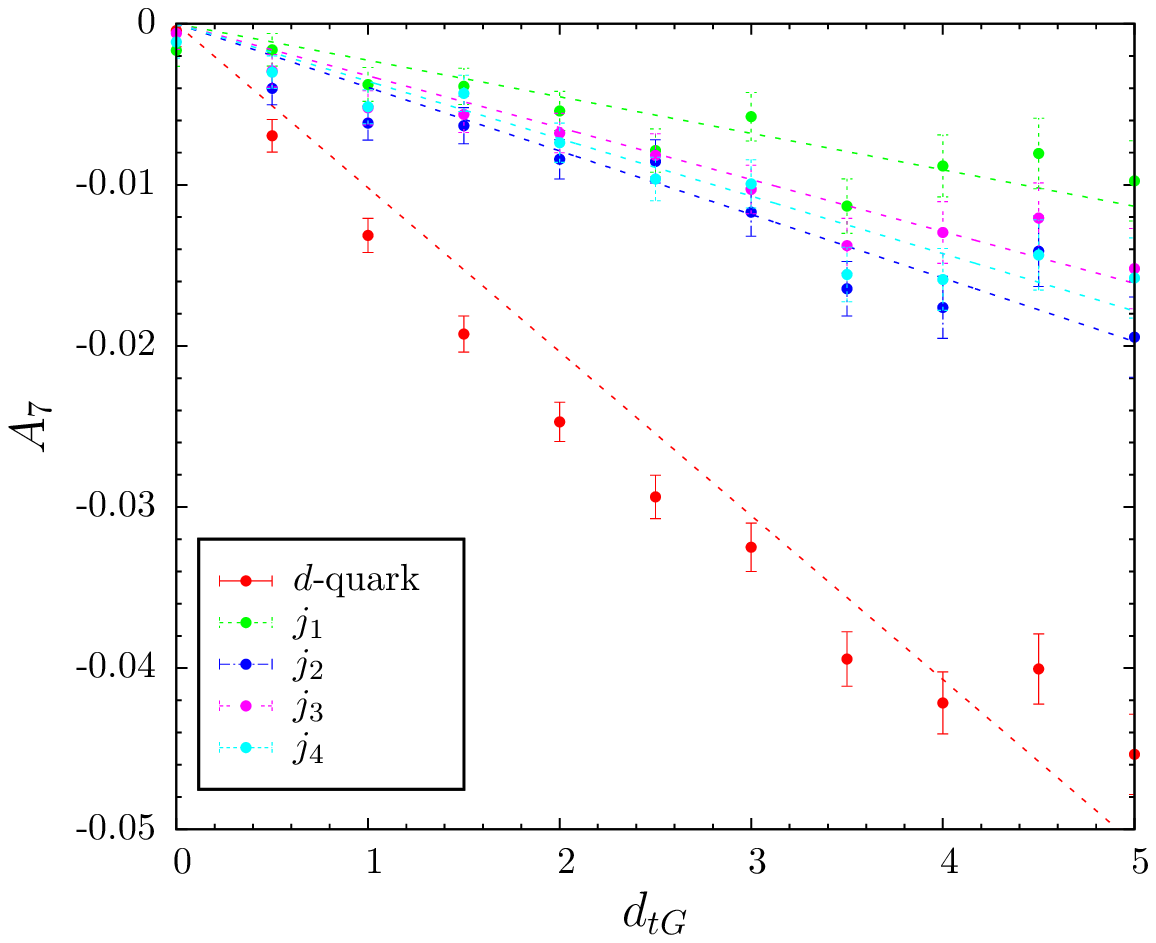} \hspace*{0.1cm}
\includegraphics[scale=0.4]{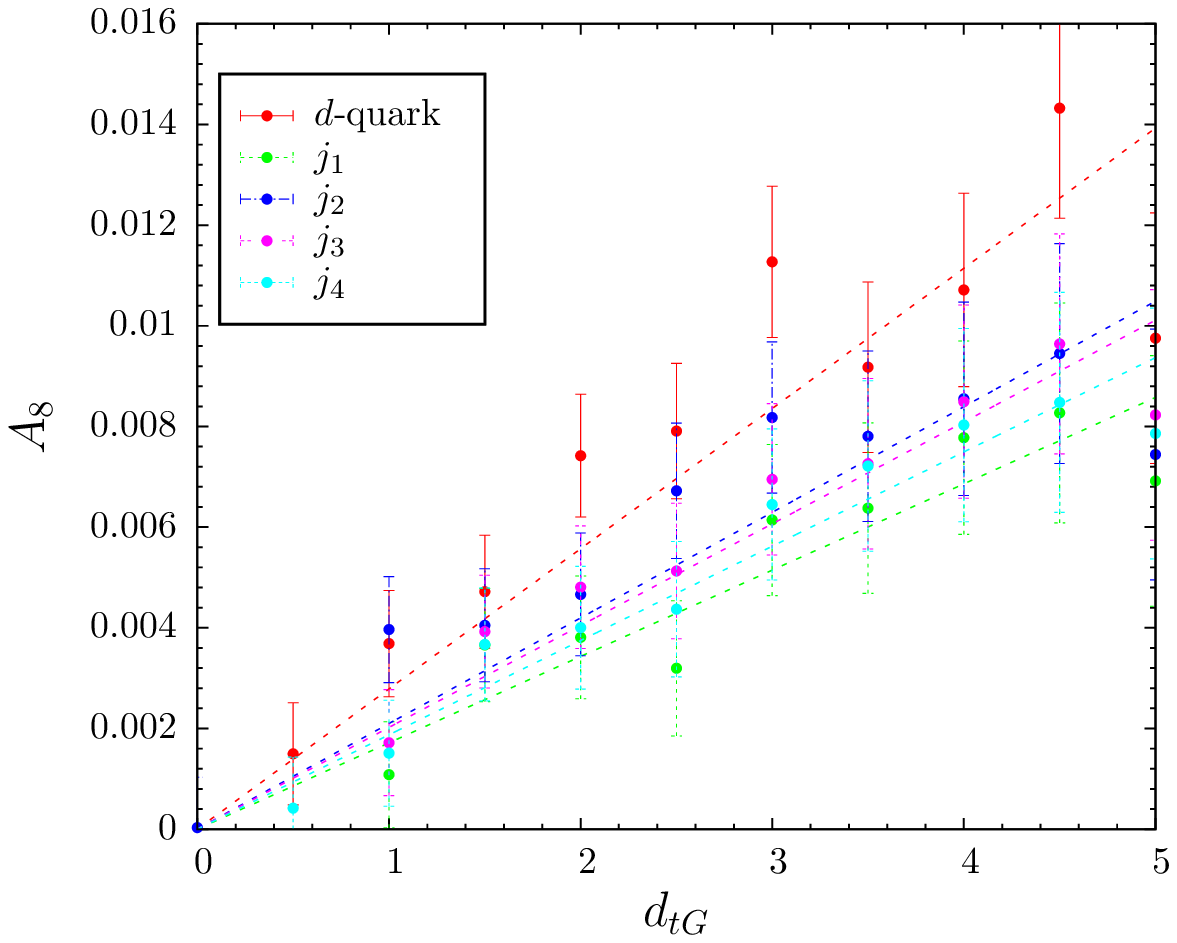} \hspace*{0.1cm}
\includegraphics[scale=0.4]{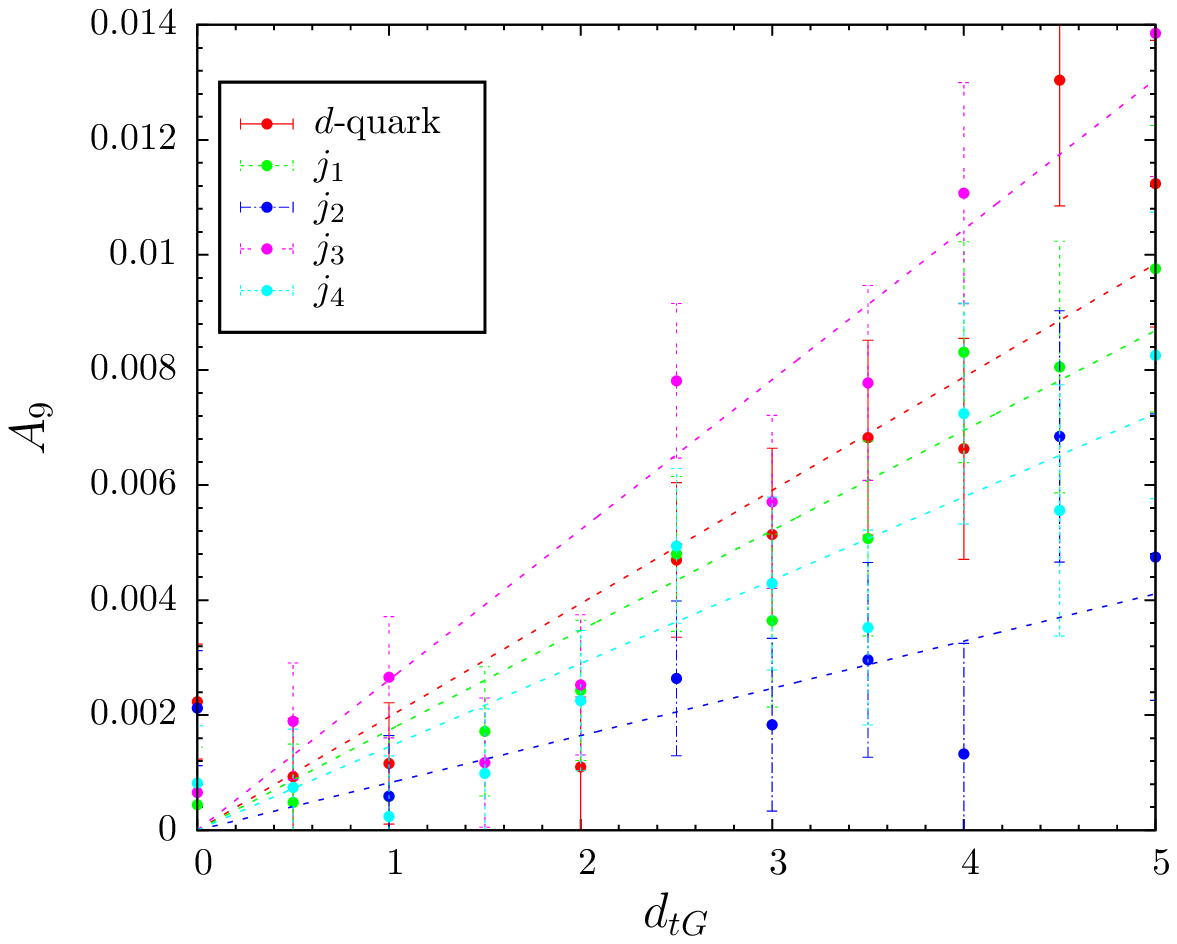} 
\includegraphics[scale=0.4]{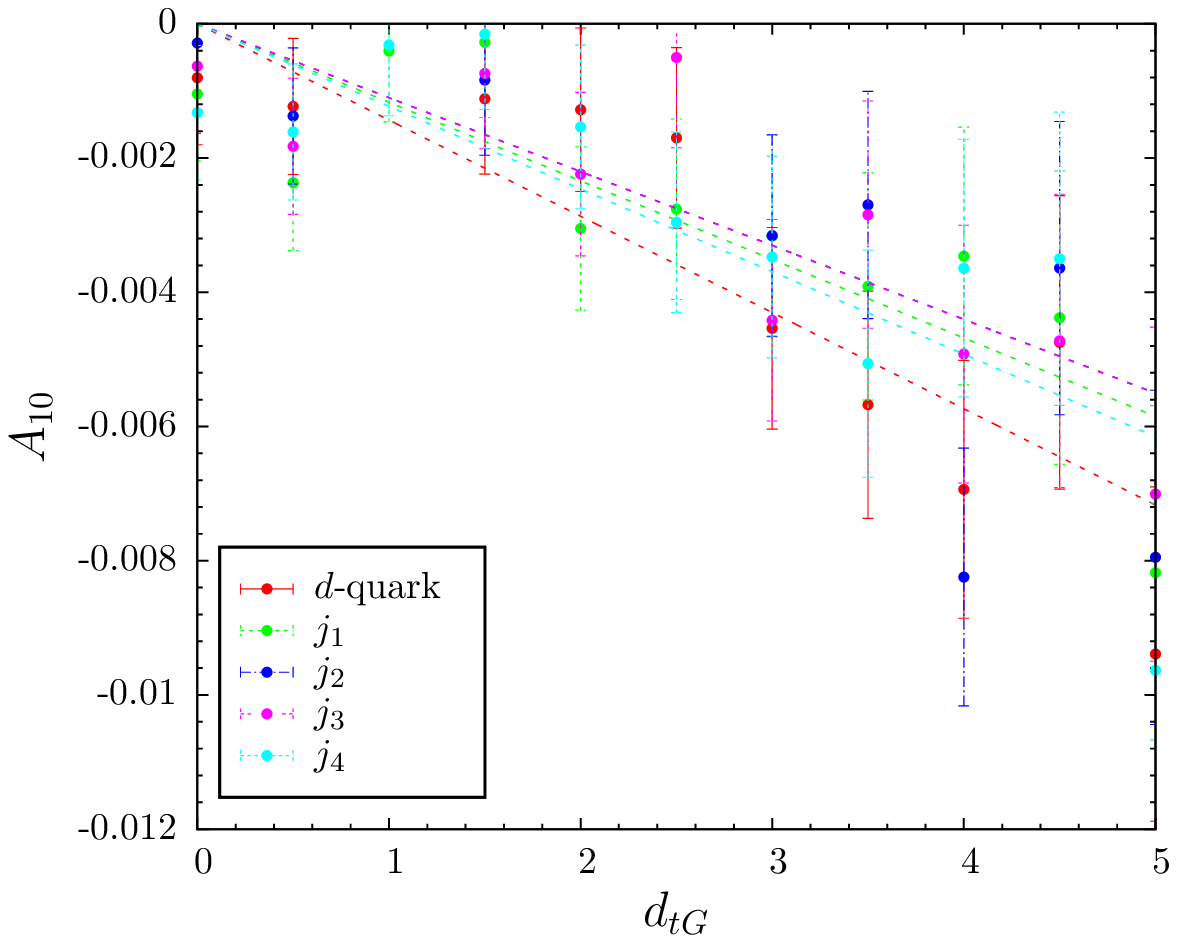} \hspace*{0.1cm}
\includegraphics[scale=0.4]{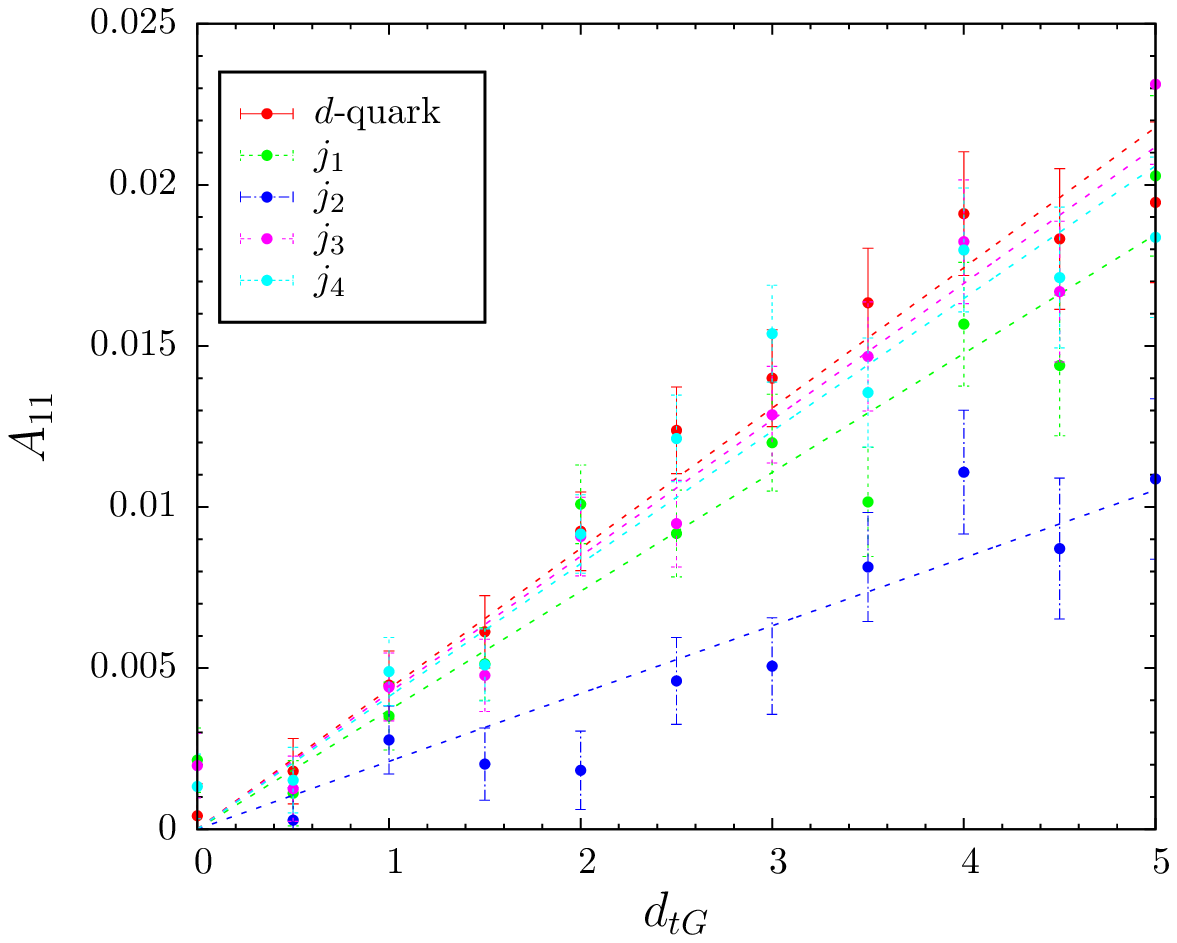} \hspace*{0.05cm}
\includegraphics[scale=0.4]{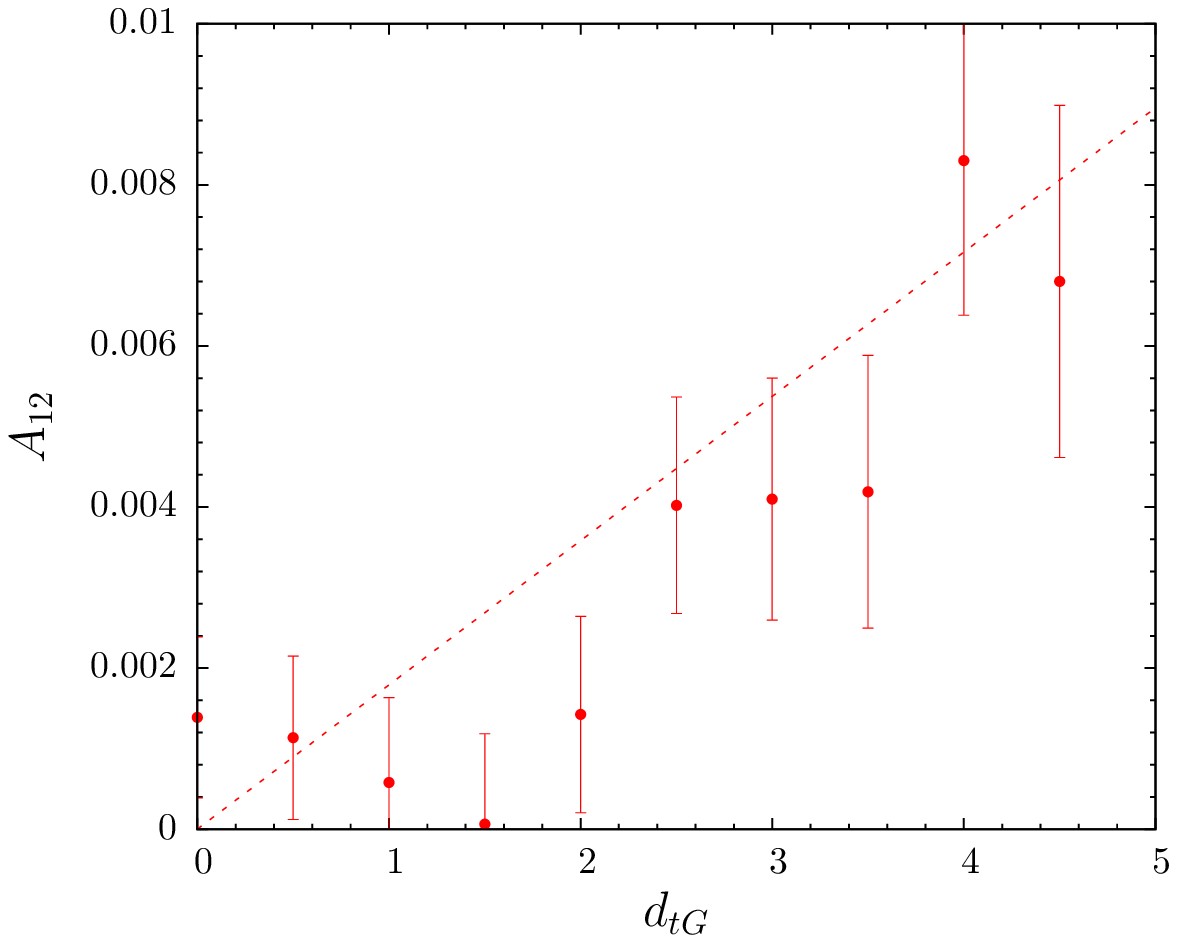} 
\includegraphics[scale=0.4]{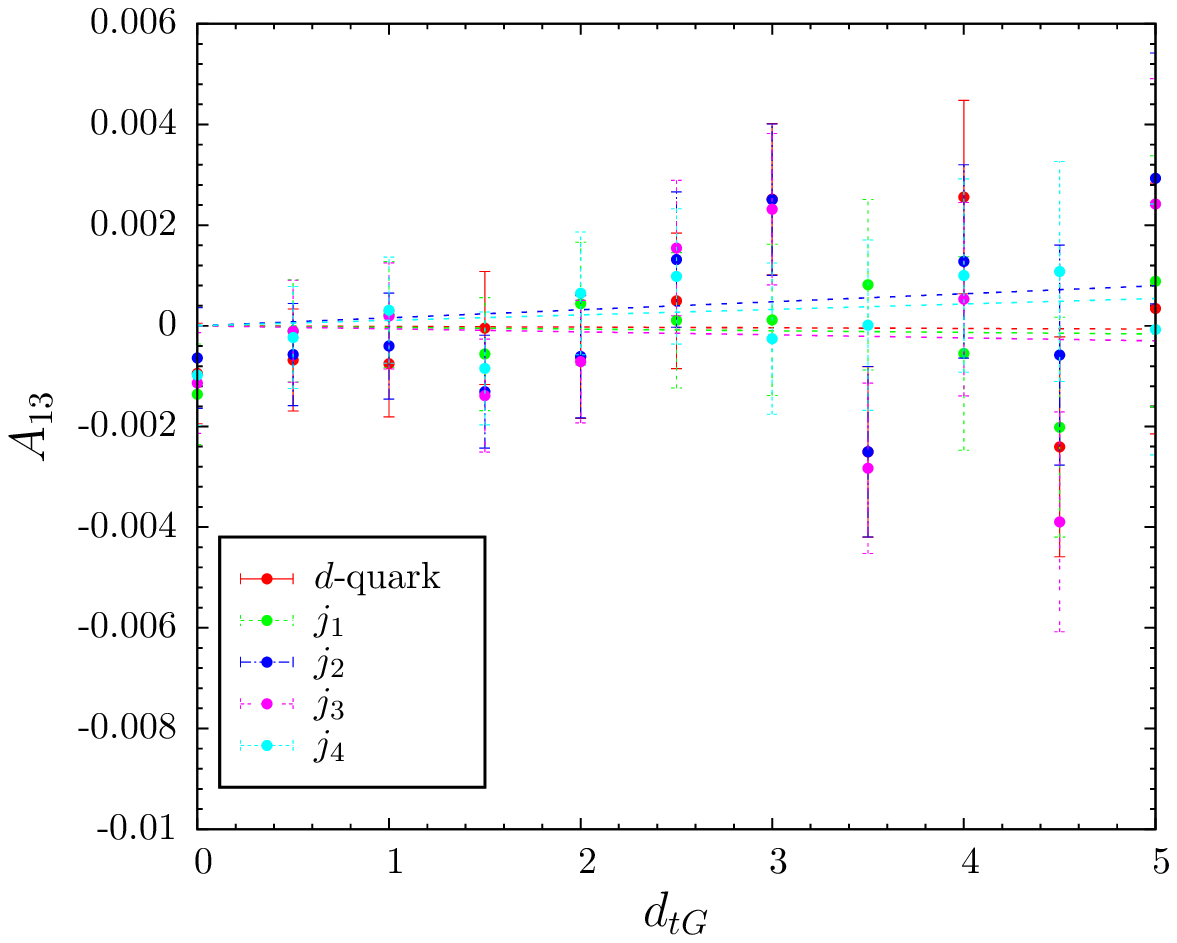}  \hspace*{0.1cm}
\includegraphics[scale=0.4]{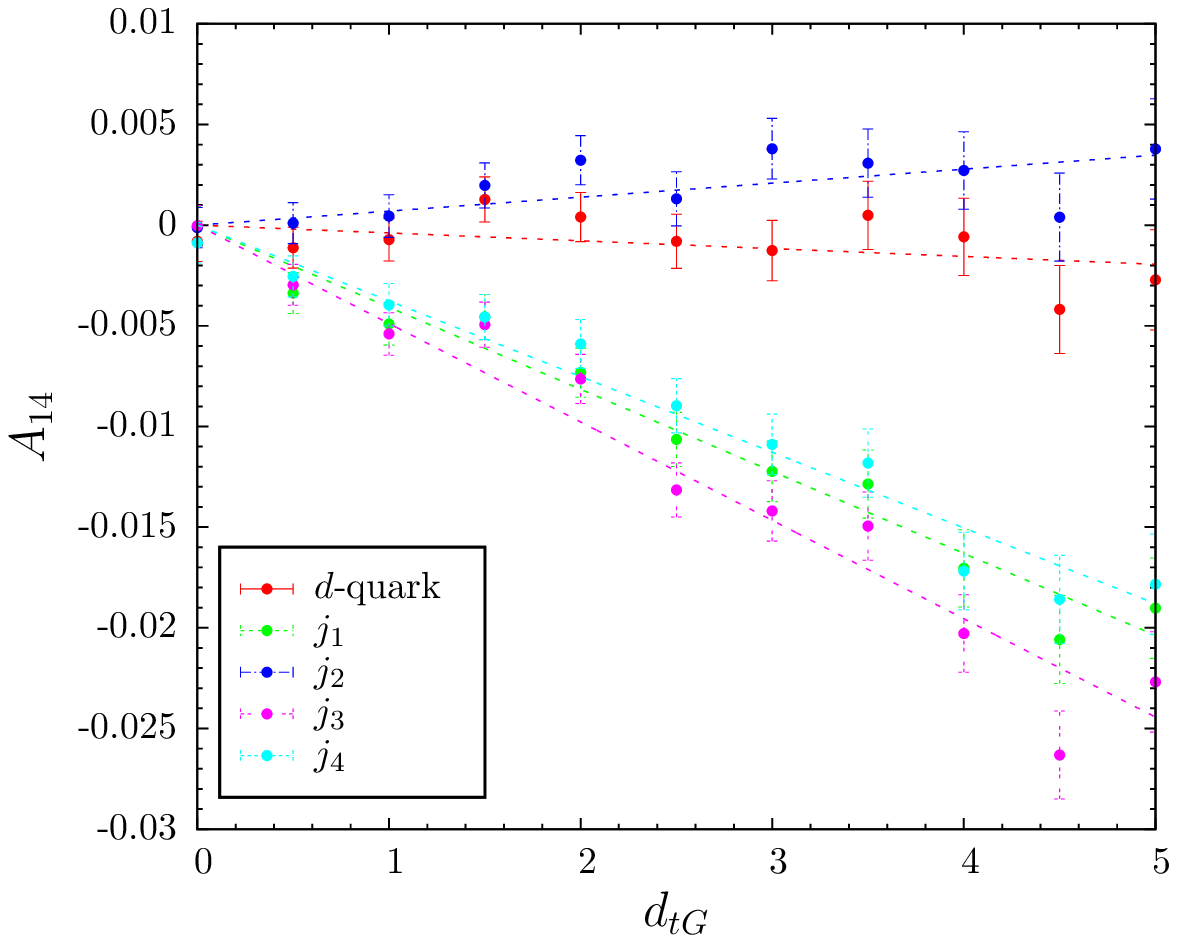}
\caption{Asymmetries calculated from samples of $10^6$ MC events with their estimated statistical error and our best linear fits. In all cases where there is a top-quark four-momentum (or antitop-quark) we use the visible momentum as defined in the text.}
\label{fig}
\end{figure}

Our numerical results are collected in Fig.~\ref{fig}. They can be summarized with fits to the 14 observables of the form
\begin{equation}
A_i=c_i\ d_{tG}\left(\frac{1{\rm ~TeV}}{\Lambda}\right)^2
\label{eqfit}
\end{equation}
with the coefficients $c_i$ tabulated in Table~\ref{tab2} . The coefficient $c_{13}$ is zero as can be seen from Figure~\ref{fig}.

\begin{table}[htp]
\vspace*{-1.0cm}
\begin{tabular}{|c|c|c|c|}
\hline 
 & ${\cal{O}}_i$ &$j$&$c_i$  \\
\hline 
\multirow{4}{1em}{1} & \multirow{4}{11em}{$q_\ell \ \ \epsilon(t,\bar{t},\ell,j)$} & 1 & -0.0094 \\
& & 2 &-0.0159 \\
& & 3 & -0.0163 \\
&  & 4 & -0.0160 \\
\hline
\multirow{1}{1em}{2}  &\multirow{1}{11em}{ $\epsilon(t,\bar{t},b,\bar{b})$} &- & -0.0160  \\
\hline 
\multirow{4}{1em}{3} & \multirow{4}{11em}{$q_\ell \ \ \epsilon(b,\bar{b},\ell,j)$} & 1 & -0.0148 \\
& & 2& -0.0157 \\ 
& & 3& -0.0198 \\
& &4& -0.0160 \\
\hline 
\multirow{4}{1em}{4} & \multirow{4}{11em}{$\epsilon(b^\ell,b^j,\ell,j)$} &  1 & -0.0041   \\ 
&&2& -0.0055  \\ 
&&3 & -0.0057 \\ 	
&&4 & -0.0048 \\ 	
\hline
\multirow{1}{1em}{5} & \multirow{1}{11em}{ $q_\ell \  q\cdot\ell \epsilon(b,\bar{b},\ell,q)$} &- & -0.0022 \\
\hline 
\multirow{4}{1em}{6} & \multirow{4}{11em}{$q_\ell \ \ \epsilon(P,b-\bar{b},\ell,j)$ }&  1 &  0.0095 \\
&&2 & 0.0120 \\
&&3& 0.0140 \\ 
&&4& 0.0117 \\ 
\hline  
\multirow{4}{1em}{7} & \multirow{4}{11em}{$q_\ell \ \ q\cdot(t-\bar{t}) \epsilon(P,q,\ell,j)$} & 1 & -0.0023 \\
&&2&-0.0039  \\
&&3 & -0.0032 \\
&&4 & -0.0036 \\
\hline 
\end{tabular}
\quad
\begin{tabular}{|c|c|c|c|}
\hline 
 & ${\cal{O}}_i$ & $j$&$c_i$  \\
\hline 
\multirow{4}{1em}{9} & \multirow{4}{11em}{$q\cdot\ell \epsilon(b+\bar{b},q,\ell,j)$} & 1 & 0.0017 \\
&&2&  0.0008 \\
&&3& 0.0026 \\
&&4& 0.0014  \\
\hline 
\multirow{4}{1em}{10} & \multirow{4}{11em}{$q\cdot(b-\bar{b})\epsilon(b,\bar{b},q,j)$}& 1 &-0.0012 \\
&&2& -0.0011 \\
&&3& -0.0011 \\
&&4&-0.0012 \\
\hline 
\multirow{4}{1em}{11} & \multirow{4}{11em}{$q_\ell \  q\cdot(b-\bar{b})\epsilon(P,q,b+\bar{b},j)$}& 1 & 0.0037 \\
&&2& 0.0021 \\
&&3& 0.0042 \\
&&4&  0.0041 \\
\hline 
\multirow{1}{1em}{12} & \multirow{1}{11em}{$q\cdot(b-\bar{b})\ \epsilon(P,q,b,\bar{b})$}& - & 0.0018\\
\hline 
\multirow{4}{1em}{14} & \multirow{4}{11em}{ $\epsilon(P,b+\bar{b},\ell,j)$} & 1 & -0.0041 \\   
&&2& 0.0007 \\
&&3& -0.0049 \\
&&4& -0.0038 \\
\hline 
\end{tabular}
\quad
\begin{tabular}{|c|c|c|c|}
\hline 
\multirow{4}{1em}{8} & \multirow{4}{18em}{$q\cdot(t-\bar{t})(P\cdot\ell\epsilon(q,b,\bar{b},j)+P\cdot j\epsilon(q,b,\bar{b},\ell))$}& 1&0.0017 \\
&&2& 0.0021 \\
&&3& 0.0020  \\
&&4&0.0019 \\
\hline
\end{tabular}
\label{tab2}
\caption{Asymmetry coefficient $c_i$ for Eq.~(\ref{eqfit}) for the four different ways to pick the jet. Note that $t$ or $\bar{t}$ denote the {\it visible} top (or antitop) momenta as defined in Eq.~(\ref{ptvis}).}
\end{table}

Several conclusions can be drawn from Table~\ref{tab2}. 
\begin{itemize}
\item The asymmetries that do not require distinguishing between a $b$ and a $\bar{b}$, $A_{9,10,12,13}$, are not very sensitive to the CEDM; in fact  they are consistent with zero within our statistical error. 
\item In view of this, and because it may not be possible to completely distinguish the $b$ and $\bar{b}$ jets at LHC, we propose ${\cal O}_4$ in which the $b$'s are classified by closeness to the lepton ($b^\ell$) or hardest jet ($b^j$) as defined by $\Delta R$. 
\item It appears that ${\cal O}_1$ has the largest sensitivity. Since the $t$ or $\bar{t}$ momentum cannot be fully reconstructed in the lepton plus jets channel we use the visible top momenta and this makes ${\cal O}_{1,2,3}$ related. For example 
\begin{eqnarray}
{\cal O}_{1}[p_{t-vis},j_4]&=&\epsilon(b+\ell^+,\bar{b}+W^-,\ell^+,W^-)-\epsilon(b+W^+,\bar{b}+\ell^-,\ell^-,W^+)\nonumber \\
&=&\epsilon(b,\bar{b},\ell^+,W^-)-\epsilon(b,\bar{b},\ell^-,W^+)=
{\cal O}_3[j_4]
\end{eqnarray}
as can be seen from simple kinematics.
\item Table~\ref{tab0} shows that the sensitivity of the lepton plus jets channel is in principle as good as that of the dimuon channel. There is substantial dilution in going from an unobservable $d$ quark to a jet. As Table~\ref{tab2} shows, however, a judicious choice of the jet to go in the asymmetry can improve sensitivity by factors of 2. The only condition in choosing this jet is that it should be CP blind: the probability should be the same in $t$ or $\bar{t}$ hadronic decay.
\item To arrive at the true sensitivity it will be necessary to simulate events at the hadron level and try different jet definitions; this task is better suited for the experimental collaborations.
\end{itemize}

\subsection{Background}

The dominant background processes discussed by the CMS and ATLAS collaborations are single top production, $W$ plus jets, $Z$ plus jets and QCD. None of these backgrounds is CP violating in the SM and hence they cannot contribute to true CP odd observables (those without the four-vectors $P$ or $q$). In these cases they can only dilute the asymmetries by resulting in a larger measured cross section. The level at which this can occur can be inferred from the cross-section measurements, and in the 7~TeV CMS lepton plus jets analysis they would be under 6.5\% \cite{Chatrchyan:2012ria}
\begin{equation}
\sigma_{t\bar{t}}=(158.1\pm 2.1({\rm stat}) \pm 10.2 ({\rm syst})\pm 3.5({\rm lum})  ){\rm ~pb}
\end{equation}
The asymmetries involving initial state momenta could be faked by unitarity phases \cite{Hayreter:2015cia}, such as those appearing at higher order in QCD \cite{Hagiwara:2007sz}. This issue has not been fully studied for top-quark pair production, but at least in this case it can be addressed by studying the asymmetries $A_{1}$ and $A_3$ for example.

\subsection{Dilution factors}

There are several experimental factors that will affect the measurement of any of these asymmetries. Among them, not being able to distinguish between $b$ and $\bar b$ jets, misidentifying the event as a top-pair event, misreconstructed objects, spatial resolution for particle momenta and so on. A simple way to parametrize these effects is with a dilution factor
\begin{eqnarray}
A_{\rm exp ~i} = \varepsilon A_i
\end{eqnarray}
where the dilution factor $\varepsilon$ will be a product of dilution factors from all the experiment effects. For example, we have identified two of them in this paper for ${\cal O}_3$:
\begin{eqnarray}
\varepsilon_{b {\rm ~vs~} \bar{b}}& \sim & 0.3 \\
\varepsilon_{{\rm not~}t\bar{t}} &\sim &  0.96
\end{eqnarray}
The first factor is estimated by comparing $A_3$ and $A_4$ but note that the experiments may do better than this. The second factor is estimated from Table~2 in Ref.~\cite{Chatrchyan:2012ria}. The remaining cross sections from the main background processes (in lepton plus jets at 7 TeV) were found to be: single top $1.17\pm 0.10$~pb; $W+{\rm jets}$ $3.35\pm 0.26$~pb; $Z +{\rm jets}$ $1.43\pm 0.29$~pb, resulting in the number quoted above.
This simple parametrization in terms of dilution factors will work as long as there is no CP violation in the background.

\section{Summary}

Recent studies have dealt with placing limits on the CEDM of the top-quark by studying deviations from the SM cross section for top-pair production at the LHC. To single out the CP violating nature of this coupling, T-odd correlations were proposed with simulations concentrating on the dimuon channel. In this paper we extended the numerical studies of T-odd correlations to the lepton plus jets channel motivated by its higher statistics and interest from the experimental collaborations. We first identified operators in lepton plus jets that correspond to operators in the dilepton channel and compared their sensitivity. Once we established the form of the operator through this comparison, we constructed observables at the parton level in terms of jet momenta. We studied the effect of different jet definitions on the sensitivity of the observables. One of the most sensitive observables we found, ${\cal O}_3$, corresponds to $T_2$ discussed in Ref.~\cite{Sjolin:2003ah} who found a similar sensitivity to the CEDM as we did. 

\begin{acknowledgments}

This research was supported in part by the DOE under Contract No. DE-SC0009974 and in part by the ARC Centre of Excellence for Particle Physics at the Terascale. 
We thank Kai-Feng Chen and Jui-Fa Tsai for many useful discussions.

\end{acknowledgments}

\appendix

\section{Summary of results}

The lepton plus jets events are generated with commands  \\ {\tt generate $p p \rightarrow t \bar{t}, (t \rightarrow b l^+ \nu), (\bar{t} \rightarrow \bar{b} j j)$ \\
add process $p p \rightarrow t \bar{t}, (\bar{t} \rightarrow \bar{b} l^- \bar{\nu}), (t \rightarrow b j j)$} \\ with the cuts of Eq.~(\ref{eqcuts}). Our numerical results are summarized in the figure below. We generate samples of $10^6$ events after cuts for each of ten different values of the coupling ${\rm Im}(d_{tG})$ and plot the asymmetry calculated from these events along with its statistical error.  We fit these points with a straight line going through the origin (as there is no asymmetry for pure SM) to obtain the numbers shown in Table~\ref{tab2}.

The notation used in writing the correlations is shown below for ${\cal O}_3$ as an example. With $p^\mu\equiv(E,p_x,p_y,p_z)$,
\begin{eqnarray}
{\cal O}_3\ =\  q_\ell\epsilon(b,\bar{b},\ell,j) &\equiv& q_\ell \epsilon_{\mu\nu\alpha\beta}p_b^\mu p_{\bar b}^\nu p_\ell^\alpha p_j^\beta
\end{eqnarray}
where for the Levi-Civit\`{a} tensor we use the convention $\epsilon_{0123}=-1$. This can be written as a determinant,
\begin{eqnarray}
{\cal O}_3\ =\  q_\ell \left| \begin{array}{cccc} p_{bx} & p_{by} & p_{bz} & E_b \\
p_{\bar b x} & p_{\bar b y} & p_{\bar b z} & E_{\bar b} \\
p_{\ell x} & p_{\ell y} & p_{\ell z} & E_\ell \\
p_{j x} & p_{j y} & p_{j z} & E_j \end{array}\right|
\end{eqnarray}
In the form given, ${\cal O}_i$ is a Lorentz scalar that can be calculated in any frame and that turns into the more familiar triple product correlation in specific frames. In this case, in the $b{\bar b}$ center of mass it becomes
\begin{eqnarray}
{\cal O}_3 &\xrightarrow[]{ (b{\bar b})_{\rm c.m.}} -\dfrac{q_\ell}{2} m_{b\bar{b}} (\vec{p}_b-\vec{p}_{\bar b})\cdot(\vec{p}_{\ell}\times\vec{p}_{j})
\end{eqnarray}

\end{document}